\documentclass[twocolumn]{aastex631}
\usepackage{amssymb}
\usepackage{amsmath}
\usepackage{gensymb}
\usepackage{graphicx}
\usepackage{xcolor}
\usepackage{natbib}

\newcommand{\beq}{\begin{equation}}
\newcommand{\eeq}{\end{equation}}

\newcommand{\Ms}{\textrm{M}_*}
\newcommand{\Msun}{\textrm{M}_\odot}
\newcommand{\kmps}{km~s$^{-1}$}
\newcommand{\MHI}{\rm{M_{H{\textsc i}}}}

\newcommand{\MB}{{\rm M_B}}

\newcommand{\hi}{H{\sc i}}
\newcommand{\himf}{H{\sc i}MF}

\newcommand{\hii}{H{\sc i} 21\,cm}

\graphicspath{{./}{figures/}}

\shorttitle{The \hi\ mass function at $z\approx1$}
\shortauthors{Chowdhury, Kanekar, and Chengalur}

\begin{document}

\title{The \hi\ Mass Function of Star-forming Galaxies at $z\approx1$}	

	\correspondingauthor{Nissim Kanekar}
	\email{nkanekar@ncra.tifr.res.in}
	
	\author{Aditya Chowdhury}
	\affil{National Centre for Radio Astrophysics, Tata Institute of Fundamental Research, Pune, India.}
	
	\author{Nissim Kanekar}
	\affil{National Centre for Radio Astrophysics, Tata Institute of Fundamental Research, Pune, India.}
	
	\author{Jayaram N. Chengalur}
	\affil{National Centre for Radio Astrophysics, Tata Institute of Fundamental Research, Pune, India.}

	
	
\begin{abstract}
We present the first estimate, based on direct H{\sc{i}}~21\,cm observations,  of the H{\sc{i}} mass function (H{\sc{i}}MF) of star-forming galaxies at $z\approx1$, obtained by combining our measurement of the scaling relation between \hi\ mass ($\rm{M_{H{\textsc{i}}}}$) and B-band luminosity ($\rm{M_B}$) of star-forming galaxies with literature estimates of the B-band luminosity function at $z\approx1$. We determined the $\rm{M_{H{\textsc{i}}}-M_B}$ relation by using the GMRT-CAT$z1$ survey of the DEEP2 fields to measure the average \hi\ mass of blue galaxies at $z=0.74-1.45$ in three separate $\rm{M_B}$ subsamples. This was done by separately stacking the H{\sc{i}}~21\,cm emission signals of the galaxies in each subsample to detect, at $(3.5-4.4)\sigma$ significance, the average H{\sc{i}}~21\,cm emission of each subsample. We find that the $\rm{M_{H{\textsc{i}}}-M_B}$ relation at $z\approx1$ is consistent with that at $z\approx0$. { We combine our estimate of the $\rm{M_{H{\textsc{i}}}-M_B}$ relation at $z\approx1$ with the B-band luminosity function at $z\approx1$ to determine the H{\sc{i}}MF at $z\approx1$. We find that the number density of galaxies with $\rm{M_{H{\textsc{i}}}>10^{10}~M_\odot}$ (higher than the knee of the local H{\sc{i}} mass function) at $z\approx1$ is a factor of $\approx4-5$ higher than that at $z\approx0$, for a wide range of assumed scatters in the $\rm{M_{H{\textsc{i}}}-M_B}$ relation. We rule out the hypothesis that the number density of galaxies with $\rm{M_{H{\textsc{i}}}>10^{10}~M_\odot}$ remains unchanged between $z \approx 1$ and $z\approx0$ at $\gtrsim99.7$\% confidence. This is the first statistically significant evidence for evolution in the H{\sc{i}}MF of galaxies from the epoch of cosmic noon.}

\end{abstract}
	
	\keywords{Galaxy evolution --- Neutral hydrogen clouds --- High-$z$ galaxies}
	
\section{Introduction}

Neutral atomic hydrogen (\hi) is the primary fuel for star formation in galaxies and a critical component of a galaxy's baryonic cycle \citep[e.g.][]{Peroux20}. Understanding the evolution of the \hi\ content of galaxies with cosmological time is thus critical for an understanding of galaxy evolution. A basic descriptor of the \hi\ content of galaxies at any epoch is the ``\hi\ mass function'' ({\hi}MF), the number density of galaxies of a given \hi\ mass as a function of the \hi\ mass \citep[e.g.][]{Bothun85,Briggs90,Briggs93,Rao93,Zwaan97,Zwaan05,Rosenberg02,Hoppmann15,Jones18}. Over the past two decades, unbiased wide-field \hii\ emission surveys have shown that a Schechter function provides a good fit to the \himf\ at $z\approx0$, with an exponential decline at masses above the ``knee'' of the mass function and a power-law dependence at low \hi\ masses \citep[e.g.][]{Zwaan05,Jones18}. At present, the most accurate measurement of the \himf\ in the local Universe is from the ALFALFA survey with the Arecibo Telescope \citep{Haynes18}, yielding a knee \hi\ mass of log($\rm M_*/M_\odot)=9.94 \pm 0.05$, a low-mass power-law slope of $\alpha=-1.25 \pm 0.1$, and a normalization of $\phi_* = (4.7 \pm 0.8) \times 10^{-3}$~Mpc$^{-3}$~dex$^{-1}$, where the errors are dominated by systematic uncertainties due to cosmic variance and the absolute flux density scale \citep{Jones18}. 

Unfortunately, the weakness of the \hii\ line has meant that little is known about the \himf\ at cosmological distances. So far, direct estimates of the \himf\ via unbiased \hii\ surveys are restricted to $z\approx0.1$ \citep{Hoppmann15,Ponomareva23}. Measuring the \himf\ at high redshifts is of critical importance to understanding galaxy evolution. Indeed, while different cosmological hydrodynamic simulations \citep[e.g. SIMBA, IllustrisTNG, EAGLE;][]{Schaye15,Pillepich19,Dave19} do reproduce the \himf\ at $z \approx 0$, the results of these simulations for the \himf\ are very different at $z \approx 1$ and $z \approx 2$ \citep[e.g.][]{Dave20}. Measurements of the \himf\ at high redshifts thus offer an avenue to distinguish between different models of galaxy evolution, and their inbuilt assumptions and subgrid physics.

The inherent weakness of the \hii\ line can be overcome by using the stacking approach, in which the \hii\ emission signals from a sample of galaxies with known spectroscopic redshifts are combined to obtain the average \hi\ mass of the sample \citep{Zwaan00,Chengalur01}. Such \hii\ stacking has been used to characterise the \hi\ properties of different galaxy populations out to $z\approx1.3$ \citep[e.g.][]{Bera19,Bera23,Bera23b,Chowdhury20,Chowdhury21,Chowdhury22a,Chowdhury22d}.  Recently, \citet{Bera22} demonstrated that \hii\ stacking can also be used to determine the \himf, by combining a measurement of the $\MHI-\MB$ relation at $z \approx 0.34$ (obtained from \hii\ stacking) with the B-band luminosity function ($\phi(\MB)$) at the same redshift. This approach, an extension of that used to obtain the early estimates of the \himf\ in the local Universe \citep[see, e.g., ][]{Briggs90,Rao93,Zwaan01} yielded the first measurement of the \himf\ at intermediate redshifts.

In this Letter, we use data from the Giant Metrewave Radio Telescope (GMRT) Cold-\hi\ AT $z\approx1$ (GMRT-CAT$z1$) survey \citep{Chowdhury22b} to determine the $\MHI-\MB$ relation for star-forming galaxies at $z\approx1$. We combine this relation with measurements of the B-band luminosity function at $z\approx1$ to obtain the first measurement of the \himf\ at $z\approx1$, at the end of the epoch of cosmic noon. 

\section{The GMRT-CAT$z$1 Survey}

The GMRT-CAT$z1$ survey used 510~hr of total time with the Band-4 receivers of the upgraded GMRT to observe three sky fields of the DEEP2 Galaxy Redshift Survey \citep{Newman13}, covering the frequency range $\approx 550-830$~MHz. This allowed us to carry out an \hii\ emission survey of galaxies at $z=0.74-1.45$, over an $\approx2$~sq.~deg area, divided into seven GMRT pointings. The observations (carried out over 3~GMRT cycles), the data analysis, the galaxy sample,  and the \hii\ emission stacking procedure are described in detail in \citet{Chowdhury22b}. We provide below, for completeness, a brief summary of the sample of galaxies and the approach used for the \hii\ stacking.

The main sample of the GMRT-CAT$z1$ survey contains 11,419 blue star-forming galaxies at $z=0.74-1.45$ with accurate spectroscopic redshifts \citep[redshift accuracy $\lesssim 62$~\kmps;][]{Newman13} in the DEEP2 DR4 catalogue, obtained after excluding (i)~red galaxies, based on the color-magnitude relation between rest-frame U-B color and absolute B-magnitude $\MB$ \citep{Willmer06}, (ii)~radio-loud active galactic nuclei based on their rest-frame 1.4~GHz luminosity \citep{Condon02}, (iii)~low-mass galaxies with $\Ms<10^9~\Msun$, and (iv)~galaxies whose \hii\ spectra were found to be affected by systematic non-Gaussian issues \citep{Chowdhury22b}. { We emphasize that the exact choices of the thresholds for the above selection criteria do not have a significant effect on our measurement of the average \hi\ mass of the full sample of galaxies \citep{Chowdhury22b}.} We further restrict the sample for the present analysis to galaxies with $\MB \leq -20$,  the B-band completeness limit of the DEEP2 survey { for blue galaxies} at $z \approx 1$ \citep{Newman13}. This yields a final sample of 10,177 blue star-forming galaxies with $\MB \leq -20$ at $z \approx 0.74-1.45$. 

For each of the three observing cycles, the GMRT-CAT$z$1 survey yielded a spectral cube covering $\approx 550-830$~MHz for each pointing that was observed in the cycle \citep{Chowdhury22b}. Each galaxy of the sample thus typically has $2-3$ observations of its \hii\ emission, obtained from the different observing cycles. For each galaxy of the sample, a subcube was extracted from its main 
spectral cube covering a spatial extent of $\pm500$~kpc around the galaxy position and $\pm1500$~\kmps\ around its redshifted \hii\ line frequency. All subcubes have a spatial resolution of 90~kpc  and a velocity resolution of 90~\kmps\ in the rest frame of the galaxy\footnote{Throughout this work, we consistently use a flat Lambda-Cold Dark Matter 737 cosmology, with $\Omega_m=0.3$, $\Omega_\Lambda = 0.7$, and $H_0 = 70$~\kmps~Mpc$^{-1}$.}. In all, we obtained 25,892 subcubes for the 10,177 galaxies in the present sample.

The spatial resolution of { 90~proper~kpc} was chosen to ensure that the average \hii\ emission signal from the full sample of galaxies in the GMRT-CAT$z1$ survey is not resolved \citep{Chowdhury22b}. { We further note that \hii\ source confusion does not significantly affect our average \hi\ mass measurements at the spatial resolution of 90~kpc \citep{Chowdhury22b}.}

\section{The $\MHI-\MB$ Relation at $z\approx1$}
\label{sec:mhi-mb}

\subsection{The average \hi\ mass for galaxy subsamples}

\begin{figure}
    \centering
    \includegraphics[width=\linewidth]{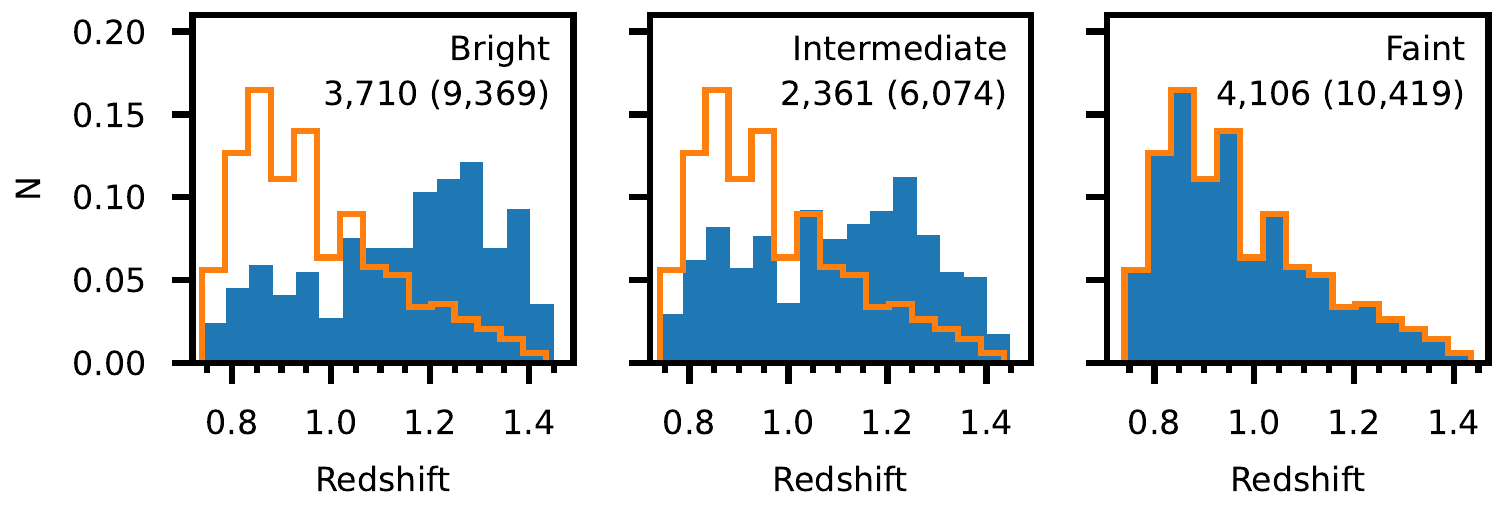}
\caption{The redshift distributions of the galaxies in the three $\MB$ subsamples, shown in blue.  The distributions have been normalised by the total number of subcubes in each subsample. The \hii\ subcubes of each $\MB$ subsample were assigned weights such that the effective redshift distribution of each subsample is identical to the redshift distribution of the faint $\MB$ subsample (orange histograms). The number of galaxies in each subsample is indicated in each panel, with the number of \hii\ subcubes shown in parentheses.}
 \label{fig:redshiftdist}
\end{figure}

To determine the $\MHI-\MB$ relation at $z\approx1$, we first divided the sample of 10,177 blue galaxies at $z=0.74-1.45$ into three $\MB$ subsamples with $\MB<-21.3$ (bright), $-21.3\le\MB<-20.9$ (intermediate), and  $-20.9\le\MB\le-20.0$ (faint). The number of galaxies and the number of \hii\ subcubes in each $\MB$ subsample are listed in Table~\ref{tab:mbStacks}, while the redshift distributions of the three $\MB$ subsamples are shown in  Figure~\ref{fig:redshiftdist}. The redshift distributions of the three subsamples are clearly different. This is because the DEEP2 Survey targetted galaxies for spectroscopy down to a magnitude limit of R$_\textrm{AB}=24.1$, resulting in a bias towards galaxies with higher B-band luminosity at higher redshifts \citep{Willmer06,Newman13}. We corrected for this bias by using weights in the \hii\ stacking such that the effective redshift distributions of the bright and intermediate $\MB$ subsamples are identical to the redshift distribution of the faint $\MB$ subsample. We separately stacked the \hii\ subcubes of the galaxies in the three subsamples, using the above weights to ensure that the redshift distributions of the three subsamples are identical.

For each $\MB$ subsample, the stacked \hii\ cube was obtained by taking a weighted mean, across all \hii\ subcubes in the sample, of the measured \hii\ luminosity density in each spatial and velocity pixel of the individual \hii\ subcubes \citep{Chowdhury22b}. We note that the weights used during stacking were only to ensure that the redshift distribution of the subsamples is identical; no additional variance-based weights were used. Any residual spectral baselines were subtracted out by fitting a second-order polynomial to the spectrum of each spatial pixel of the stacked \hii\ cube, excluding velocity channels covering the central $\pm250$~\kmps. The RMS noise on each stacked \hii\ cube was obtained via Monte Carlo simulations. In each realisation of the Monte Carlo runs, the redshift of each DEEP2 galaxy in the subsample was shifted by a value in the range $\pm1500$~\kmps, drawn randomly from a uniform distribution. The velocity-shifted \hii\ subcubes were then stacked, using a procedure identical to that followed for the ``true'' \hii\ stack, to obtain a realisation of the stacked \hii\ cube. This procedure was repeated to produce $10^4$ stacked \hii\ cubes, from which we measured the RMS noise on each spatial and velocity pixel.

\begin{figure*}
\centering
\includegraphics[width=\linewidth]{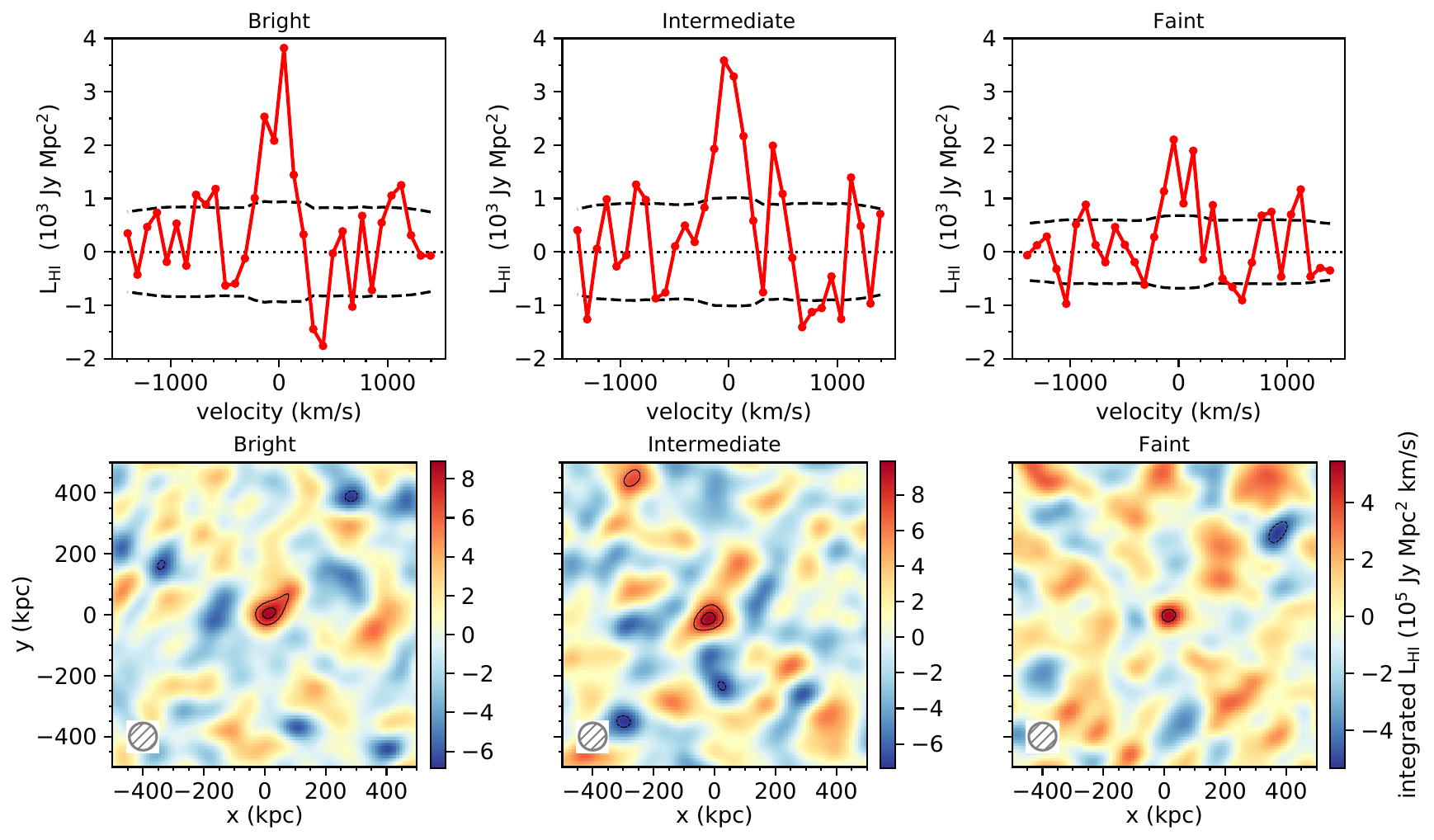}
\caption{The average \hii\ emission signal from star-forming galaxies in different $\MB$ subsamples at $z \approx 1$. The top panels show the stacked \hii\ emission spectra, at a velocity resolution of 90~\kmps, for galaxies in the three $\MB$ subsamples, bright (left panel), intermediate (middle panel), and faint (right panel). The dashed curve in each panel shows the $\pm1\sigma$ error on the stacked \hii\ spectrum. 
The bottom panels show the average \hii\ emission images of the same galaxies in the three $\MB$ subsamples, obtained by integrating the \hii\ emission over the central velocity channels ($=180$~\kmps\ to $+180$~\kmps) of the stacked spectral cubes. The circle at the bottom left of each panel shows the 90-kpc spatial resolution of the images. The contour levels are at $-3.0\sigma$ (dashed), $+3.0\sigma$, and $+4.0\sigma$ significance; {note that there are no $-3\sigma$ features in the images}. The average \hii\ emission signals from all three subsamples are seen to be clearly detected in both the spectra and the images.}
\label{fig:mbStacks}
\end{figure*}

Figure~\ref{fig:mbStacks} shows the stacked \hii\ emission images and \hii\ spectra for the three $\MB$ subsamples. We obtain detections, with $\approx 3.5-4.4\sigma$ statistical significance, of the average \hii\ emission signal from the galaxies in each of the three subsamples. For each $\MB$ subsample, the average \hi\ mass of the constituent galaxies was obtained from the stacked \hii\ cube using the following procedure: (i)~the central velocity channels of the cube were integrated to produce an image of the \hii\ emission signal, (ii)~a spectrum was obtained from the stacked \hii\ cube at the location of the peak luminosity density of the above \hii\ image, (iii)~a contiguous range of velocity channels with \hii\ emission detected at $\ge1.5\sigma$ significance were selected to obtain a measurement of the average velocity-integrated \hii\ line luminosity ($\int {\rm L_{H\textsc{i}} \ dV}$), in units of ${\rm Jy~Mpc^2}$~\kmps, and (iv)~the average velocity-integrated \hii\ line luminosity was converted to the average \hi\ mass of the sample via the relation $\MHI=1.86 \times 10^4 \times \int {\rm L_{H\textsc{i}} \  dV}$, in units of $\Msun$. For all three $\MB$ subsamples, the velocity interval with \hii\ emission detected at $\ge 1.5\sigma$ significance over contiguous velocity channels was found to be [$-180$~\kmps, $+180$~\kmps]. The average \hi\ masses of the galaxies in the three $\MB$ subsamples are listed in Table~\ref{tab:mbStacks}.


\begin{table*}
\centering
\begin{tabular}{|l|c|c|c|}
\hline
\hline
    \,  & Bright & Intermediate & Faint \\
    \hline
    $\MB$~Range & $[-23.8,-21.3]$  &  $[-21.3,-20.9]$  & $[-20.9,-20.0]$ \\
    \hline
     Number of \hii\ Subcubes & $9,369$ & $6,074$ & $10,419$\\
     \hline
    Number of Galaxies & $3,710$ & $2,361$ & $4,106$\\
     \hline
    Average Redshift   & 0.97 & 0.97 & 0.97 \\
    \hline
    Average $\MB$  & $-21.73$ & $-21.09$   & $-20.50$  \\
       \hline
    Average $\Ms$ ($\times 10^{9}\ \Msun$) & $23.3$ & $9.9$   & $4.8$  \\
   \hline
    Average \hi\ Mass  ($\times 10^{9}\ \Msun$) & $16.6\pm3.9$ & $18.5\pm4.2$ & $10.2\pm2.8$ \\

     \hline
\end{tabular}
\caption{The average properties of the galaxies in the three $\MB$ subsamples. For each $\MB$ subsample, the rows are (1)~the $\MB$ range, (2)~the number of \hii\ subcubes, (3)~the number of galaxies, (4)~the average redshift, (5)~the average $\MB$, (6)~the average stellar mass, and (7)~the average \hi\ mass, measured from the stacked \hii\ emission spectra of Figure~\ref{fig:mbStacks}. Note that the average redshifts, \hi\ masses, $\MB$ values, and stellar masses are all weighted averages, with the weights chosen to ensure an identical redshift distribution for the three $\MB$ subsamples.}
\label{tab:mbStacks}
\end{table*}

\subsection{The $\MHI-\MB$ relation at $z \approx 1$}

\begin{figure}
    \centering
    \includegraphics[width=\linewidth]{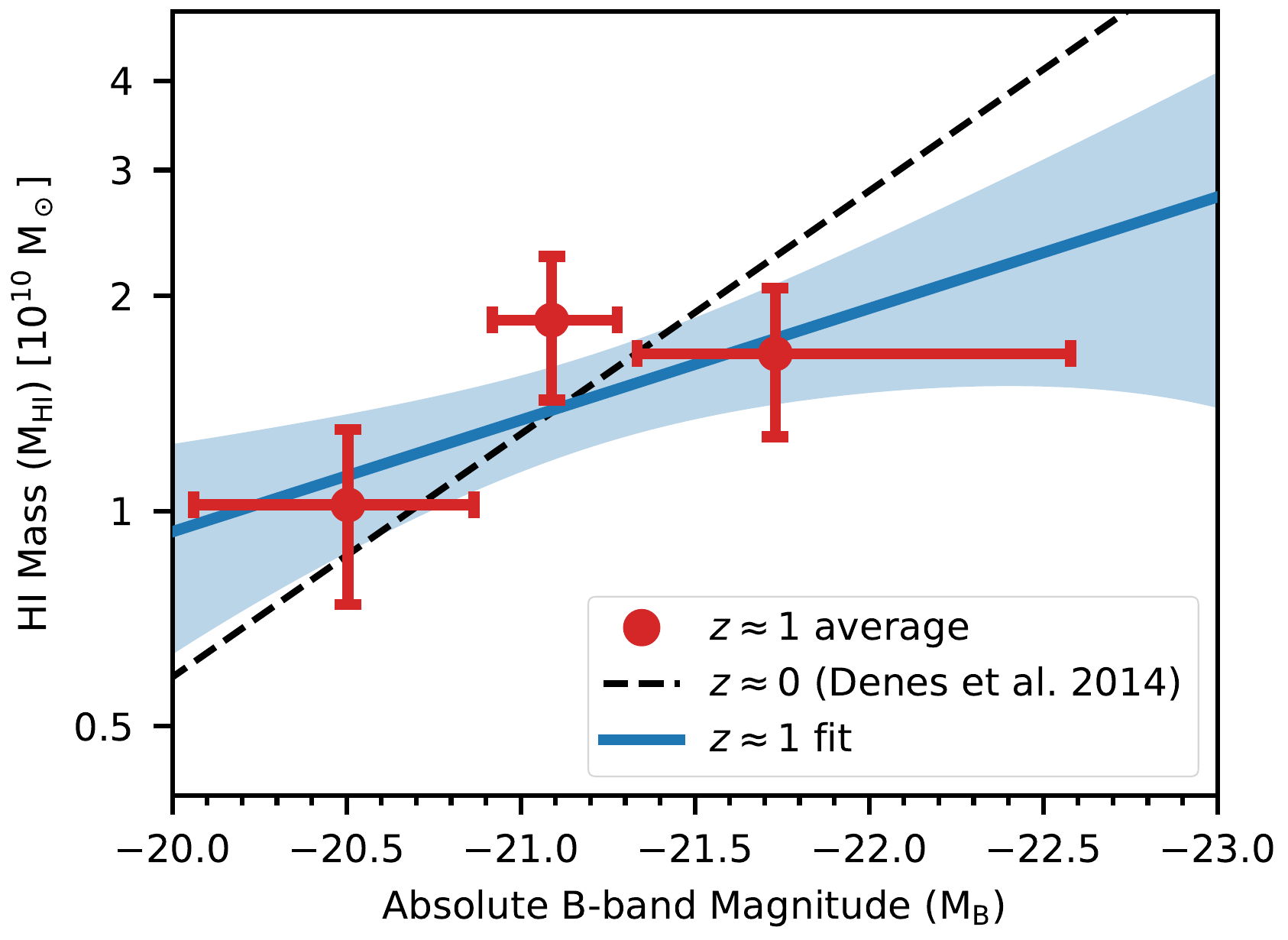}
    \caption{The $\MHI - \MB$ relation at $z\approx1$. The red points show our measurements of the average \hi\ mass in the three $\MB$ subsamples. The blue curve and blue shaded region show, respectively, the best fit to the measurements and the $1\sigma$ confidence interval. The black dashed line indicates the ``mean'' $\MHI - \MB$ relation at $z\approx 0$ \citep{Denes14}. It is clear that our measurements of the average \hi\ mass in the three $\MB$ subsamples are consistent with the $\MHI-\MB$ relation at $z \approx 0$, indicating that the relation has not evolved between $z\approx1$ and $z\approx0$. We emphasise that the figure shows the ``mean" $\MHI-\MB$ relation at both $z\approx1$ and $z\approx0$ (see Section~\ref{sec:mhi-mb}): we have used the measured scatter of 0.26~dex in the $\MHI-\MB$ relation at $z\approx0$ \citep{Denes14} and Equation~\ref{eqn:med-mean} to obtain the mean $\MHI-\MB$ relation at $z\approx0$.}
    \label{fig:MHI_MB}
\end{figure}

We fitted a power-law relation to our measurements of the average \hi\ mass of star-forming galaxies at $z\approx1$ in the three $\MB$ subsamples, following the procedures of Appendix~\ref{appndx:fit}, to obtain the $\MHI-\MB$ relation. The best-fit $\MHI-\MB$ relation at $z\approx1$ is
\begin{equation}
\label{eqn:MHI_MB_full}
\begin{split}
   \log \left[\MHI/\Msun\right]= &(-0.156\pm0.105)  \times (\MB+21) \\
   &+(10.127\pm0.068) \,.
\end{split}
\end{equation}

In the local Universe, \hi\ scaling relations such as the $\MHI-\MB$ relation are obtained from measurements of the \hi\ mass in individual galaxies, fitting a relation to $\langle \log \MHI\rangle$ as a function of $\MB$ \citep[e.g.][]{Denes14}. The distribution of \hi\ masses in each such galaxy subsample is typically log-normal \citep[e.g.][]{Catinella18}. In such cases, $\langle \log \MHI\rangle$ is equal to the logarithm of the median of the \hi\ masses of the subsample. The scaling relations obtained in the local Universe from measurements of individual \hi\ masses are thus ``median'' scaling relations. However, in stacking analysis, scaling relations such as the $\MHI-\MB$ relation of Equation~\ref{eqn:MHI_MB_full} are based on measurements of the average \hi\ mass and hence of  $\log \langle \MHI\rangle$ in multiple galaxy subsamples. The scaling relations directly obtained from stacking analyses are thus ``mean'' scaling relations and are generally different from the median scaling relations obtained from measurements of individual \hi\ masses \citep{Bera22,Bera23,Chowdhury22d}.

The above difference between the median scaling relations obtained at $z\approx0$ from \hii\ studies of individual galaxies and the mean scaling relations obtained via \hii\ stacking analyses at high redshifts  must be accounted for when comparing scaling relations at different redshifts. 
Assuming that the distribution of \hi\ masses is log-normal with a logarithmic scatter $\sigma$, one obtains 
\begin{equation}
\langle \log \MHI \rangle=\log\langle\MHI\rangle - (\ln 10/2) \sigma^2 \,.
\label{eqn:med-mean}
\end{equation}
This implies that the intercept of the mean scaling relation ($\alpha_\mathrm{mean}$) obtained from a stacking analysis is related to that of the median scaling relation ($\alpha_\mathrm{med}$) obtained from individual detections via $\alpha_\mathrm{med}=\alpha_\mathrm{mean}- (\ln 10/2)\sigma^2$. The slopes of the two relations are the same.

Figure~\ref{fig:MHI_MB} shows our measurements of $\langle\MHI\rangle$ in the three $\MB$\ subsamples and the fit of Equation~\ref{eqn:MHI_MB_full} to the measurements. For comparison, the figure also shows the mean $\MHI - \MB$ relation for late-type galaxies at $z \approx 0$ \citep{Denes14}, where we have used the measured scatter of 0.26~dex in the median local relation \citep{Denes14} to obtain the mean relation \citep{Bera22,Chowdhury22d}. The figure shows that our measurements of $\langle \MHI \rangle$ in the three $\MB$ subsamples at $z \approx 1$ are remarkably consistent with the $\MHI-\MB$ relation at $z\approx0$. While the $\MHI-\MB$ relation appears slightly flatter at $z\approx1$ than at $z \approx 0$, the slope of $-0.156\pm0.105$ at $z \approx 1$ is formally consistent (within $2\sigma$ significance) with the slope of $-0.34 \pm 0.01$ at $z\approx0$ \citep{Denes14}. 

Finally, a measurement of the \himf\ from the $\MHI-\MB$ relation requires the median relation \citep{Bera22}. We assume that the scatter of the $\MHI-\MB$ relation is the same at $z=1$ as at $z=0$, i.e. we assume that the scatter in the relation at $z\approx1$ to also be 0.26~dex; {\bf the effect of this assumption is discussed later}. This allows us to obtain the median $\MHI-\MB$ relation at $z \approx 1$ from the mean $\MHI-\MB$ relation of Equation~\ref{eqn:MHI_MB_full}; this yields
\begin{equation}
\label{eqn:MHI_MB_median}
\begin{split}
   \log \left[\MHI/\Msun\right]=    & (-0.156\pm0.105)  \times (\MB+21) \\ 
   &+(10.049\pm0.068) \,.
\end{split}
\end{equation}

\subsection{Systematic effects}

{We note that our measurement of the $\MHI-\MB$ relation at $z\approx1$ could be affected by incompleteness in the galaxy sample of the DEEP2 survey. We provide below a systematic exploration of possible biases in our measurement of the $\MHI-\MB$ relation due to such sample incompleteness. 

First, the DEEP2 Galaxy Redshift survey targeted galaxies for spectroscopy down to a limiting magnitude R$_\textrm{AB}=24.1$. The survey selected galaxies from a photometric catalogue with a limiting magnitude that is 1.5~mag. fainter than the threshold for DEEP2 spectroscopy and yielded a redshift success rate that is approximately independent of magnitude and colour \citep{Newman13}. As such, photometric incompleteness of the DEEP2 galaxy sample is unlikely to be an issue for our results.

Next, the R-band selection of the DEEP2 survey yields a rest-frame B-band completeness that is a function of both redshift and colour: for a fixed $\MB$, the survey is biased toward galaxies with increasingly bluer rest-frame (U-B) colour at higher redshifts \citep{Newman13}. In our current study, we have restricted our sample for \hii\ stacking to galaxies with $\MB\le-20$, the B-band magnitude at which the DEEP2 survey is complete to blue star-forming galaxies at $z\lesssim1$ \citep{Willmer06,Newman13}. This implies that the galaxy sample at $z\gtrsim1$ used for the \hii\ stacking could be biased towards bluer galaxies, which could affect our average \hi\ mass measurements. However, \citet{Chowdhury22a} find that the DEEP2 galaxies in our sample at $z>1$ have average stellar masses and average star formation rates (SFRs) consistent with the star-forming main-sequence relation derived from highly complete photometric surveys at these redshifts. This indicates that the average properties of the DEEP2 galaxies in our sample are representative of the general main-sequence galaxy population over $0.74 \lesssim z \lesssim 1.45$.

Finally, we note that our choice of weights for galaxies in the three subsamples is based on the distribution of galaxies in the faintest subsample and is hence biased toward the galaxies at $z\lesssim1$ (see Figure~\ref{fig:redshiftdist}) for which the DEEP2 sample is complete at $\MB\le-20$ \citep{Newman13}. Indeed, galaxies at $z>1$ make up $\approx33\%$ of each $\MB$ subsample, after the redshift-based weighting, while galaxies at $z>1.2$, for which the effects of incompleteness would be the highest, make up only $\approx10\%$ of each subsample. Our average \hi\ mass measurements in the three subsamples are thus likely to be robust to any potential effects of incompleteness in the higher-redshift bins.  Overall, we conclude that our measurement of the $\MHI-\MB$ relation at $z\approx1$ is likely to be robust to the effects of the DEEP2 selection criteria.}

\section{Determining the \hi\ Mass Function at $z\approx1$}
\label{sec:obs:himf}

While modern measurements of the \himf\ in the local Universe have relied on wide-area optically-unbiased \hii\ emission surveys \citep[e.g.][]{Zwaan05,Jones18}, the early estimates of the \himf\ at $z\approx0$ were based on combining the $\MHI-\MB$ relation with measurements of the B-band luminosity function $\phi(\MB)$ \citep[e.g.][]{Briggs90,Rao93,Zwaan01}. These authors performed a simple transformation of variable in $\phi(\MB)$, substituting $\MB$ with $\MHI$ via the observed $\MHI-\MB$ relation, to obtain $\phi(\MHI)$. However, \citet{Bera22} found that this straightforward approach, which ignores the scatter in the $\MHI-\MB$ relation, results in an underestimation of the \himf\ at the high-mass end. \citet{Bera22} further demonstrated that combining the $\MHI-\MB$ relation of \citet{Denes14} with the local B-band luminosity function, and incorporating the measured scatter of the local $\MHI-\MB$ relation \citep{Denes14}, yields an \himf\ at $z \approx 0$ in excellent agreement with that obtained from the ALFALFA survey \citep{Jones18}. {In passing, we note that it may also be possible to derive the \himf\ via other scaling relations such as the $\MHI-\Ms$ relation. However, the measured scatter of the local $\MHI-\MB$ relation  \citep[0.26~dex;][]{Denes14} is far lower than that of other \hi\ scaling relations at $z\approx0$ \citep[e.g. 0.4~dex for the $\MHI-\Ms$ relation; ][]{Catinella18}, making it a better choice for the purpose of estimating the \himf\ at high redshifts. }

To determine the \himf\ at $z \approx 1$, we need an estimate of the rest-frame B-band luminosity function of blue star-forming galaxies at this redshift. For this, we used the Schechter function fit to the B-band luminosity function, $\phi_\textrm{B}(\MB)$, of blue galaxies, obtained from the ALHAMBRA survey \citep{LopezSanjuan17}, to estimate the number density of galaxies at a given $\MB$. \citet{LopezSanjuan17} find that the redshift evolution of the ``knee'' ($\MB^{*}$) and the normalisation ($\phi^*$) of the B-band luminosity function over $z\approx0.2-1$ can be described, respectively, by $\MB^{*}=(-21.00\pm0.03)+(z-0.5)\times(-1.03\pm0.08)$ and $\log\left[\phi^{*} (10^{-3} {\rm Mpc}^{-3} {\rm mag}^{-1})\right]=(-2.51\pm0.03)+(z-0.5)\times(-0.01\pm0.03)$. Further, these authors assume that the slope of the luminosity function ($\alpha$) does not evolve with redshift, to measure $\alpha=1.29\pm0.02$.  We used the above relations for $\MB^{*}$ and $\phi^*$ to estimate the B-band luminosity function of star-forming galaxies at $z=0.97$, the mean redshift of our three $\MB$ subsamples (see Table~\ref{tab:mbStacks}). 

{In passing, we emphasize that the above errors on the parameters of the B-band luminosity function from the ALHAMBRA survey include an estimate of systematic uncertainties such as survey incompleteness, cosmic variance, etc. We take into account the above uncertainties while estimating the uncertainty in our measurement of the \himf\ at $z\approx1$. Further, the B-band luminosity function of the ALHAMBRA survey is in excellent agreement with multiple independent measurements of the B-band luminosity function at similar redshifts, including from the DEEP2 survey \citep{LopezSanjuan17}.}

\citet{Bera22} used a Monte Carlo approach to determine the \himf\ at $z \approx 0.35$, combining their measurements of the $\MHI-\MB$ relation at $z \approx 0.35$ with the rest-frame B-band luminosity function at this redshift, and assuming that the scatter in the $\MHI-\MB$ relation at $z \approx 0.35$ is the same as that at $z \approx 0$. { We follow a slightly different, but equivalent, convolution-based approach, to combine the B-band luminosity function  $\phi(\MB)$ at $z\approx1$ from the ALHAMBRA survey \citep{LopezSanjuan17} with our measurement of the median $\MHI-\MB$ relation at $z\approx1$ (Equation~\ref{eqn:MHI_MB_median}), incorporating the effect of the scatter in the $\MHI-\MB$ relation, to obtain the \himf\ at $z\approx1$. The procedure is detailed in Appendix~\ref{appndx}.}\footnote{In passing, we note that very similar results are obtained for the \himf\ on using the Monte Carlo approach followed by \citet{Bera22}.} We again assume that the scatter in the $\MHI-\MB$ relation at $z\approx1$ is 0.26~dex, as measured in the local Universe \citep{Denes14}.

 \begin{figure*}
     \centering
     \includegraphics[width=\linewidth]{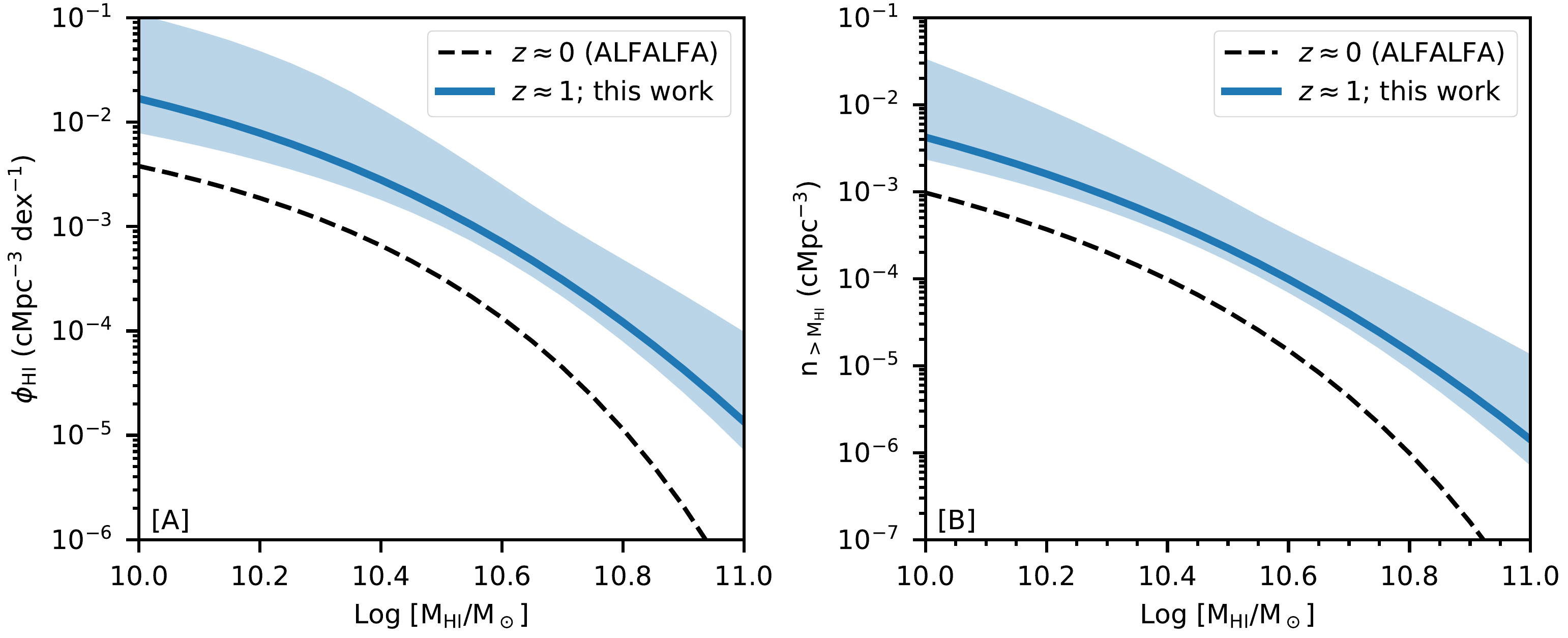}
     \caption{[A] The \himf\ of star-forming galaxies at $z\approx1$. The blue curve shows our measurement of the \himf\ at $z\approx1$, obtained by combining the GMRT-CAT$z1$ measurement of the $\MHI-\MB$ relation with the B-band luminosity function at $z \approx 1$ from the ALHAMBRA survey \citep{LopezSanjuan17}; the shaded region shows the $68\%$ confidence interval on the \himf. [B]~The number density of galaxies at $z \approx 1$ with \hi\ mass greater than $\MHI$ ($n_{>\MHI}$), obtained by integrating the \himf\ of Panel~[A] from $\MHI$ to $\infty$, is shown in blue, with the $68\%$ confidence interval indicated by the blue shaded region. The dashed curve in both panels shows the same quantities at $z\approx0$ from \citet{Jones18}. The figure shows that the number density of galaxies with $\MHI>10^{10}~\Msun$ is far larger at $z\approx1$ than at $z\approx0$. }
     \label{fig:himf}
 \end{figure*}

 \begin{figure}
     \centering
     \includegraphics[width=\linewidth]{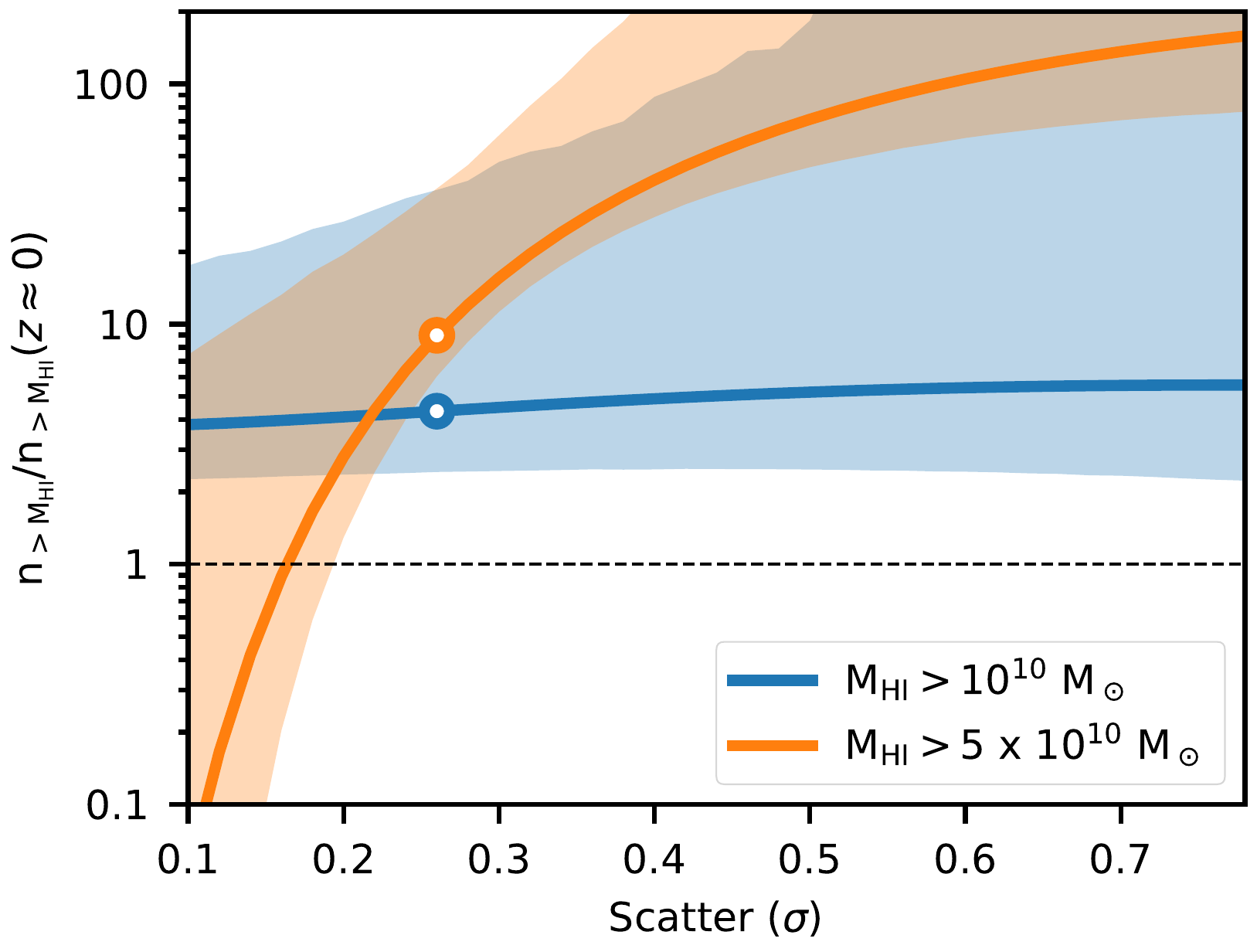}
     \caption{The ratio of the number density of galaxies at $z \approx 1$ to that at $z\approx0$ as a function of the assumed scatter in the $\MHI-\MB$ relation at $z \approx 1$, for galaxies with \hi\ masses  $> 10^{10}~\Msun$ (blue curve) and  $> 5\times10^{10}~\Msun$ (orange curve). For each curve, the shaded region indicates the $\approx68~\%$ confidence interval. The open circles indicate the ratio for an assumed scatter of $0.26$~dex, the value measured in the local Universe \citep{Denes14}. The figure shows that the number density of galaxies with $\MHI>10^{10}~\Msun$ at $z \approx 1$ is a factor of $\approx4-5$ higher than that at $z\approx0$, independent of the assumed scatter in the $\MHI-\MB$ relation. However, the same ratio for galaxies with $\MHI>5\times10^{10}~\Msun$ is seen clearly to be sensitive to the assumed scatter in the relation.}
     \label{fig:scattervariation}
 \end{figure}

\section{Results and Discussion}
\label{sec:results}

Our measurement of the \himf\ for star-forming galaxies at $z\approx1$ is shown in Figure~\ref{fig:himf}[A], with the $1\sigma$ uncertainty on the \himf\ indicated by the shaded region. Overlaid, for comparison, is the \himf\ at $z\approx0$ from the ALFALFA survey \citep{Jones18}. It is clear from the figure that the number density of high-mass galaxies, with $\MHI>10^{10}~\Msun$ {\citep[i.e. higher than the knee of the local \hi\ mass function; ][]{Jones18}}, is systematically higher at $z\approx1$ than at $z\approx0$. This can be seen more clearly in Figure~\ref{fig:himf}[B] which plots the number density of galaxies with an \hi\ mass greater than $\MHI$ ($n_{>\MHI}$), obtained by integrating the \himf\ of Figure~\ref{fig:himf}[A] from $\MHI$ to $\infty$. 
{The figure shows that the number density of galaxies with $\MHI>10^{10}~\Msun$ at $z\approx1$ is a factor of $4.3^{+29.5}_{-1.9}$ (asymmetric errors, from the $68\%$ confidence interval\footnote{We note that the upper error range is significantly larger than the lower error range, here and in the later estimates. This is because the \hi\ mass function depends non-linearly on the values of the parameters of the $\MHI-\MB$ relation, and the quoted uncertainties include the full formal errors on the slope of this relation, from Equation~\ref{eqn:MHI_MB_median}. Specifically, the number density of galaxies with large \hi\ masses is inversely related to the slope of the $\MHI-\MB$ relation, with the value rising rapidly as one approaches a slope of 0. Our current measurement of the slope of the $\MHI-\MB$ relation is consistent with 0 at $\approx 1.5\sigma$ significance, implying that the upper bound on the number density of massive galaxies at $z\approx1$ is not tightly constrained.}) higher than that in the local Universe. We find that the hypothesis that the number density of galaxies with $\MHI>10^{10}~\Msun$ does not evolve between $z\approx 1$ and $z\approx 0$ is ruled out at $\approx 99.7\%$ confidence (equivalent to  $\approx 3\sigma$ statistical significance, for Gaussian statistics). Further, considering even more massive galaxies, with $\MHI>5\times10^{10}~\Msun$ at $z\approx1$, the number density at $z \approx 1$ is a factor of $9.0^{+27.2}_{-2.9}$ higher than that in the local Universe; the hypothesis that the number density of such galaxies does not evolve from $z \approx 1$ to $z \approx 0$ is also ruled out at $\approx 99.7$\% confidence.
}

{We note that the above results are based on the assumption that the scatter in the $\MHI-\MB$ relation at $z \approx 1$ is the same as that at $z=0$. Figure~\ref{fig:scattervariation} shows the ratio of the number density of galaxies at $z \approx 1$ to that at $z \approx 0$, for  $\MHI >10^{10}~\Msun$ (blue curve) and $\MHI > 5 \times 10^{10}~\Msun$ (orange curve), as a function of the assumed scatter in the $\MHI-\MB$ relation at $z \approx 1$. The figure clearly shows that changing the assumed value of the scatter has little effect on the number density of galaxies for \hi\ masses $> 10^{10} \, \Msun$. For example, for an assumed scatter of $0.13$~dex, half that in the local Universe, the number density of galaxies with $\MHI >10^{10}~\Msun$ at $z\approx1$ is a factor of $3.9^{+14.7}_{-1.6}$ higher than that in the local Universe. Conversely, for an assumed scatter of $0.52$~dex, twice that in the local Universe, the number density of galaxies with $\MHI >10^{10} \, \Msun$ at $z\approx1$ is a factor of $5^{+247}_{-3}$ higher than that in the local Universe. In both cases, the hypothesis that the number density of galaxies with $\MHI > 10^{10}\, \Msun$ does not evolve from $z \approx 1$ to $z \approx 0$ is ruled out at  $\approx99.7\%$ confidence. Thus, for a wide range of values of the scatter in the $\MHI-\MB$ relation, the number density of galaxies with $\MHI >10^{10} \, \Msun$  at $z \approx 1$ is $\approx 4-5$ times higher than that in the local Universe. Our result that the number density of galaxies with \hi\ masses higher than the knee of the local \hi\ mass function is significantly higher at $z \approx 1$ than at $z \approx 0$ thus appears to be a robust one.
}

{The situation is different for the highest-mass galaxies, with $\MHI > 5 \times 10^{10} \, \Msun$, where the inferred number density at $z \approx 1$ does depend critically on the assumed scatter (see the orange curve in Figure~\ref{fig:scattervariation}). If the scatter in the $\MHI - \MB$ relation is significantly lower than the local value of 0.26~dex, then the number density of the most massive galaxies could be similar to that in the local Universe. Conversely, if the scatter in the relation is higher than in the local Universe, the number density of the most massive galaxies at $z \approx 1$ would be $\gg 10$ times higher than that at $z \approx 0$. 
This is an important issue for searches for individual detections of \hii\ emission in galaxies, which would be most sensitive to such massive galaxies. Deeper \hii\ emission surveys in the future should yield a direct estimate of the scatter of the $\MHI-\MB$ relation at $z \approx 1$ by  measuring the median $\MHI$ relation, via a median stacking of the \hii\ emission of galaxies in each $\MB$ subsample \citep[e.g.][]{Bera22},
}

 {We also emphasise that the GMRT-CAT$z1$ survey covers a large sky area of $\approx 2$ sq. degrees, corresponding to a sky volume of $10^7$ comoving~Mpc$^3$ over the redshift interval $z=0.74-1.45$. \citet{Chowdhury22b} estimate that the effect of cosmic variance on measured galaxy properties over the entire redshift interval is significantly lower than $10\%$. This is far lower than the statistical uncertainty on our measurement of the $\MHI-\MB$ relation at $z\approx1$. We thus do not expect the main conclusions of this paper to be affected by cosmic variance.}

 It is interesting to note that \citet{Bera22} find that the number density of galaxies at $z\approx0.35$ with high \hi\ masses is \emph{lower} than that in the local Universe, the opposite trend to that obtained here. A possible explanation for this apparent contradiction is that the high SFR of galaxies at $z\approx 1$ combined with the low net gas accretion rate at this redshift \citep{Chowdhury23} led to a decline in the \hi\ mass of massive galaxies from $z\approx1$ to $z\approx0.35$. However, this trend may have later reversed, over the last $\approx 4$~Gyr: \citet{Bera23b} find that the net gas accretion rate is comparable to the SFR from $z \approx 0.35$ to $z \approx 0$, which could yield an increased number density of galaxies with a high \hi\ mass at $z\approx0$ compared to that at $z\approx0.35$. Conversely, as emphasised by \citet{Bera22}, their estimate of the \himf\ at $z\approx0.35$ is based on a measurement of the $\MHI-\MB$ relation over a small cosmic volume, implying that their measurement could be affected by cosmic variance.  Wide-area measurements of the \himf\ at intermediate redshifts ($z\approx0.5$) are critical to achieving a more comprehensive understanding of the detailed evolution of the \himf\ over the past 8~Gyr.

 Finally, Figure~3 of \citet{Dave20} shows the expected \himf\ at $z \approx 1$ for the TNG100, EAGLE, EAGLE-Recal, and SIMBA hydrodynamical simulations. It is clear from this figure that TNG100, EAGLE, and EAGLE-Recal all yield a lower number density of high \hi-mass galaxies at $z \approx 1$, contrary to the results obtained in this work. SIMBA thus appears to be the only current cosmological hydrodynamical simulation that produces a higher number density of high \hi-mass galaxies at $z \approx 1$, in rough agreement with the measurement of the \himf\ at $z \approx 1$ presented here. However, even the SIMBA results for the number density of high \hi-mass galaxies at $z \approx 1$ are a factor of a few lower than the values obtained here at the highest \hi\ masses, $\gtrsim 10^{10.7}\, M_\odot$.

 \section{Summary}
 
 In this Letter, we have  determined the $\MHI-\MB$ relation for blue star-forming galaxies at $z\approx1$, by using the GMRT-CAT$z$1 survey and \hii\ stacking to measure the average \hi\ masses of galaxies in three independent $\MB$ subsamples. Our measurement of the $\MHI-\MB$ relation is consistent with the power-law relation at $z\approx0$ \citep{Denes14}, in both slope and amplitude. We used our $\MHI-\MB$ relation, with a scatter assumed to be the same as that at $z \approx 0$, along with the B-band luminosity function of blue galaxies at $z \approx 1$ from the ALHAMBRA survey to obtain the first estimate of the \hi\ mass function at $z\approx1$. We find statistically significant evidence that the number density of galaxies with high \hi\ masses is higher at $z \approx 1$ than in the local Universe. {Specifically, the number density of galaxies with $\MHI> 10^{10}~\Msun$ (i.e. higher than the knee of the local \hi\ mass function) at $z\approx1$ is a factor of $\approx 4-5$ higher than that at $z\approx0$, for a wide range of values of the assumed scatter in the $\MHI-\MB$ relation. We rule out, at $\gtrsim 99.7$\% confidence, the hypothesis that the number density of galaxies with $\MHI > 10^{10} \, \Msun$ at $z \approx 1$ is the same as that at $z \approx 0$. This is the first clear evidence for a change in the \himf\ at $z\lesssim1$. For even higher \hi\ masses, $\MHI > 5\times 10^{10}\, \Msun$, the number density at $z \approx 1$ is $\approx 9$ times higher than that at $z\approx0$, assuming that the scatter in the $\MHI-\MB$ relation at $z \approx 1$ is the same as that at $z \approx 0$. The high inferred number density of galaxies at $z\approx1$ with large \hi\ reservoirs implies that it may be possible for deep \hii\ emission surveys with today's radio telescopes to obtain detections of \hii\ emission in individual galaxies at $z\approx1$, and thus obtain even more direct constraints on the \himf\ toward the end of the epoch of cosmic noon.
 
 }

 \begin{acknowledgments}
	We thank the anonymous referee for comments and suggestions that have improved this paper. We thank the staff of the GMRT who have made these observations possible. The GMRT is run by the National Centre for Radio Astrophysics of the Tata Institute of Fundamental Research. NK acknowledges support from the Department of Science and Technology via a Swarnajayanti Fellowship (DST/SJF/PSA-01/2012-13). AC, NK, $\&$ JNC also acknowledge the Department of Atomic Energy for funding support, under project 12-R\&D-TFR-5.02-0700. NK also acknowledges many discussions on 21cm stacking and related issues with Balpreet Kaur and Apurba Bera that have contributed to this paper.
\end{acknowledgments}
    \software{Astropy \citep{astropy:2013, astropy:2018, astropy:2022}}
      
    \bibliography{bibliography.bib}

\begin{thebibliography}{}
\expandafter\ifx\csname natexlab\endcsname\relax\def\natexlab#1{#1}\fi
\providecommand{\url}[1]{\href{#1}{#1}}
\providecommand{\dodoi}[1]{doi:~\href{http://doi.org/#1}{\nolinkurl{#1}}}
\providecommand{\doeprint}[1]{\href{http://ascl.net/#1}{\nolinkurl{http://ascl.net/#1}}}
\providecommand{\doarXiv}[1]{\href{https://arxiv.org/abs/#1}{\nolinkurl{https://arxiv.org/abs/#1}}}

\bibitem[{{Astropy Collaboration} {et~al.}(2013){Astropy Collaboration},
  {Robitaille}, {Tollerud}, {Greenfield}, {Droettboom}, {Bray}, {Aldcroft},
  {Davis}, {Ginsburg}, {Price-Whelan}, {Kerzendorf}, {Conley}, {Crighton},
  {Barbary}, {Muna}, {Ferguson}, {Grollier}, {Parikh}, {Nair}, {Unther},
  {Deil}, {Woillez}, {Conseil}, {Kramer}, {Turner}, {Singer}, {Fox}, {Weaver},
  {Zabalza}, {Edwards}, {Azalee Bostroem}, {Burke}, {Casey}, {Crawford},
  {Dencheva}, {Ely}, {Jenness}, {Labrie}, {Lim}, {Pierfederici}, {Pontzen},
  {Ptak}, {Refsdal}, {Servillat}, \& {Streicher}}]{astropy:2013}
{Astropy Collaboration}, {Robitaille}, T.~P., {Tollerud}, E.~J., {et~al.} 2013,
  \aap, 558, A33, \dodoi{10.1051/0004-6361/201322068}

\bibitem[{{Astropy Collaboration} {et~al.}(2018){Astropy Collaboration},
  {Price-Whelan}, {Sip{\H{o}}cz}, {G{\"u}nther}, {Lim}, {Crawford}, {Conseil},
  {Shupe}, {Craig}, {Dencheva}, {Ginsburg}, {Vand erPlas}, {Bradley},
  {P{\'e}rez-Su{\'a}rez}, {de Val-Borro}, {Aldcroft}, {Cruz}, {Robitaille},
  {Tollerud}, {Ardelean}, {Babej}, {Bach}, {Bachetti}, {Bakanov}, {Bamford},
  {Barentsen}, {Barmby}, {Baumbach}, {Berry}, {Biscani}, {Boquien}, {Bostroem},
  {Bouma}, {Brammer}, {Bray}, {Breytenbach}, {Buddelmeijer}, {Burke},
  {Calderone}, {Cano Rodr{\'\i}guez}, {Cara}, {Cardoso}, {Cheedella}, {Copin},
  {Corrales}, {Crichton}, {D'Avella}, {Deil}, {Depagne}, {Dietrich}, {Donath},
  {Droettboom}, {Earl}, {Erben}, {Fabbro}, {Ferreira}, {Finethy}, {Fox},
  {Garrison}, {Gibbons}, {Goldstein}, {Gommers}, {Greco}, {Greenfield},
  {Groener}, {Grollier}, {Hagen}, {Hirst}, {Homeier}, {Horton}, {Hosseinzadeh},
  {Hu}, {Hunkeler}, {Ivezi{\'c}}, {Jain}, {Jenness}, {Kanarek}, {Kendrew},
  {Kern}, {Kerzendorf}, {Khvalko}, {King}, {Kirkby}, {Kulkarni}, {Kumar},
  {Lee}, {Lenz}, {Littlefair}, {Ma}, {Macleod}, {Mastropietro}, {McCully},
  {Montagnac}, {Morris}, {Mueller}, {Mumford}, {Muna}, {Murphy}, {Nelson},
  {Nguyen}, {Ninan}, {N{\"o}the}, {Ogaz}, {Oh}, {Parejko}, {Parley}, {Pascual},
  {Patil}, {Patil}, {Plunkett}, {Prochaska}, {Rastogi}, {Reddy Janga},
  {Sabater}, {Sakurikar}, {Seifert}, {Sherbert}, {Sherwood-Taylor}, {Shih},
  {Sick}, {Silbiger}, {Singanamalla}, {Singer}, {Sladen}, {Sooley},
  {Sornarajah}, {Streicher}, {Teuben}, {Thomas}, {Tremblay}, {Turner},
  {Terr{\'o}n}, {van Kerkwijk}, {de la Vega}, {Watkins}, {Weaver}, {Whitmore},
  {Woillez}, {Zabalza}, \& {Astropy Contributors}}]{astropy:2018}
{Astropy Collaboration}, {Price-Whelan}, A.~M., {Sip{\H{o}}cz}, B.~M., {et~al.}
  2018, \aj, 156, 123, \dodoi{10.3847/1538-3881/aabc4f}

\bibitem[{{Astropy Collaboration} {et~al.}(2022){Astropy Collaboration},
  {Price-Whelan}, {Lim}, {Earl}, {Starkman}, {Bradley}, {Shupe}, {Patil},
  {Corrales}, {Brasseur}, {N{\"o}the}, {Donath}, {Tollerud}, {Morris},
  {Ginsburg}, {Vaher}, {Weaver}, {Tocknell}, {Jamieson}, {van Kerkwijk},
  {Robitaille}, {Merry}, {Bachetti}, {G{\"u}nther}, {Aldcroft},
  {Alvarado-Montes}, {Archibald}, {B{\'o}di}, {Bapat}, {Barentsen},
  {Baz{\'a}n}, {Biswas}, {Boquien}, {Burke}, {Cara}, {Cara}, {Conroy},
  {Conseil}, {Craig}, {Cross}, {Cruz}, {D'Eugenio}, {Dencheva}, {Devillepoix},
  {Dietrich}, {Eigenbrot}, {Erben}, {Ferreira}, {Foreman-Mackey}, {Fox},
  {Freij}, {Garg}, {Geda}, {Glattly}, {Gondhalekar}, {Gordon}, {Grant},
  {Greenfield}, {Groener}, {Guest}, {Gurovich}, {Handberg}, {Hart},
  {Hatfield-Dodds}, {Homeier}, {Hosseinzadeh}, {Jenness}, {Jones}, {Joseph},
  {Kalmbach}, {Karamehmetoglu}, {Ka{\l}uszy{\'n}ski}, {Kelley}, {Kern},
  {Kerzendorf}, {Koch}, {Kulumani}, {Lee}, {Ly}, {Ma}, {MacBride}, {Maljaars},
  {Muna}, {Murphy}, {Norman}, {O'Steen}, {Oman}, {Pacifici}, {Pascual},
  {Pascual-Granado}, {Patil}, {Perren}, {Pickering}, {Rastogi}, {Roulston},
  {Ryan}, {Rykoff}, {Sabater}, {Sakurikar}, {Salgado}, {Sanghi}, {Saunders},
  {Savchenko}, {Schwardt}, {Seifert-Eckert}, {Shih}, {Jain}, {Shukla}, {Sick},
  {Simpson}, {Singanamalla}, {Singer}, {Singhal}, {Sinha}, {Sip{\H{o}}cz},
  {Spitler}, {Stansby}, {Streicher}, {{\v{S}}umak}, {Swinbank}, {Taranu},
  {Tewary}, {Tremblay}, {Val-Borro}, {Van Kooten}, {Vasovi{\'c}}, {Verma}, {de
  Miranda Cardoso}, {Williams}, {Wilson}, {Winkel}, {Wood-Vasey}, {Xue},
  {Yoachim}, {Zhang}, {Zonca}, \& {Astropy Project
  Contributors}}]{astropy:2022}
{Astropy Collaboration}, {Price-Whelan}, A.~M., {Lim}, P.~L., {et~al.} 2022,
  \apj, 935, 167, \dodoi{10.3847/1538-4357/ac7c74}

\bibitem[{{Bera} {et~al.}(2019){Bera}, {Kanekar}, {Chengalur}, \&
  {Bagla}}]{Bera19}
{Bera}, A., {Kanekar}, N., {Chengalur}, J.~N., \& {Bagla}, J.~S. 2019, \apjl,
  882, L7, \dodoi{10.3847/2041-8213/ab3656}

\bibitem[{{Bera} {et~al.}(2022){Bera}, {Kanekar}, {Chengalur}, \&
  {Bagla}}]{Bera22}
---. 2022, \apjl, 940, L10, \dodoi{10.3847/2041-8213/ac9d32}

\bibitem[{{Bera} {et~al.}(2023{\natexlab{a}}){Bera}, {Kanekar}, {Chengalur}, \&
  {Bagla}}]{Bera23}
---. 2023{\natexlab{a}}, \apjl, 950, L18, \dodoi{10.3847/2041-8213/acd0b3}

\bibitem[{{Bera} {et~al.}(2023{\natexlab{b}}){Bera}, {Kanekar}, {Chengalur}, \&
  {Bagla}}]{Bera23b}
---. 2023{\natexlab{b}}, \apjl, 956, L15, \dodoi{10.3847/2041-8213/acf71a}

\bibitem[{{Bothun}(1985)}]{Bothun85}
{Bothun}, G.~D. 1985, \aj, 90, 1982, \dodoi{10.1086/113901}

\bibitem[{{Briggs}(1990)}]{Briggs90}
{Briggs}, F.~H. 1990, \aj, 100, 999, \dodoi{10.1086/115573}

\bibitem[{{Briggs} \& {Rao}(1993)}]{Briggs93}
{Briggs}, F.~H., \& {Rao}, S. 1993, \apj, 417, 494, \dodoi{10.1086/173328}

\bibitem[{{Catinella} {et~al.}(2018){Catinella}, {Saintonge}, {Janowiecki},
  {Cortese}, {Dav{\'e}}, {Lemonias}, {Cooper}, {Schiminovich}, {Hummels}, \&
  {Fabello}}]{Catinella18}
{Catinella}, B., {Saintonge}, A., {Janowiecki}, S., {et~al.} 2018, \mnras, 476,
  875, \dodoi{10.1093/mnras/sty089}

\bibitem[{{Chengalur} {et~al.}(2001){Chengalur}, {Braun}, \&
  {Wieringa}}]{Chengalur01}
{Chengalur}, J.~N., {Braun}, R., \& {Wieringa}, M. 2001, \aap, 372, 768,
  \dodoi{10.1051/0004-6361:20010547}

\bibitem[{{Chowdhury} {et~al.}(2022{\natexlab{a}}){Chowdhury}, {Kanekar}, \&
  {Chengalur}}]{Chowdhury22a}
{Chowdhury}, A., {Kanekar}, N., \& {Chengalur}, J.~N. 2022{\natexlab{a}},
  \apjl, 931, L34, \dodoi{10.3847/2041-8213/ac6de7}

\bibitem[{{Chowdhury} {et~al.}(2022{\natexlab{b}}){Chowdhury}, {Kanekar}, \&
  {Chengalur}}]{Chowdhury22d}
---. 2022{\natexlab{b}}, \apjl, 941, L6, \dodoi{10.3847/2041-8213/ac9d8a}

\bibitem[{{Chowdhury} {et~al.}(2022{\natexlab{c}}){Chowdhury}, {Kanekar}, \&
  {Chengalur}}]{Chowdhury22b}
---. 2022{\natexlab{c}}, \apj, 937, 103, \dodoi{10.3847/1538-4357/ac7d52}

\bibitem[{{Chowdhury} {et~al.}(2023){Chowdhury}, {Kanekar}, \&
  {Chengalur}}]{Chowdhury23}
---. 2023, \apjl, 958, L29, \dodoi{10.3847/2041-8213/ad08c4}

\bibitem[{{Chowdhury} {et~al.}(2020){Chowdhury}, {Kanekar}, {Chengalur},
  {Sethi}, \& {Dwarakanath}}]{Chowdhury20}
{Chowdhury}, A., {Kanekar}, N., {Chengalur}, J.~N., {Sethi}, S., \&
  {Dwarakanath}, K.~S. 2020, \nat, 586, 369, \dodoi{10.1038/s41586-020-2794-7}

\bibitem[{{Chowdhury} {et~al.}(2021){Chowdhury}, {Kanekar}, {Das},
  {Dwarakanath}, \& {Sethi}}]{Chowdhury21}
{Chowdhury}, A., {Kanekar}, N., {Das}, B., {Dwarakanath}, K.~S., \& {Sethi}, S.
  2021, \apjl, 913, L24, \dodoi{10.3847/2041-8213/abfcc7}

\bibitem[{{Condon} {et~al.}(2002){Condon}, {Cotton}, \& {Broderick}}]{Condon02}
{Condon}, J.~J., {Cotton}, W.~D., \& {Broderick}, J.~J. 2002, \aj, 124, 675,
  \dodoi{10.1086/341650}

\bibitem[{{Dav{\'e}} {et~al.}(2019){Dav{\'e}}, {Angl{\'e}s-Alc{\'a}zar},
  {Narayanan}, {Li}, {Rafieferantsoa}, \& {Appleby}}]{Dave19}
{Dav{\'e}}, R., {Angl{\'e}s-Alc{\'a}zar}, D., {Narayanan}, D., {et~al.} 2019,
  \mnras, 486, 2827, \dodoi{10.1093/mnras/stz937}

\bibitem[{{Dav{\'e}} {et~al.}(2020){Dav{\'e}}, {Crain}, {Stevens}, {Narayanan},
  {Saintonge}, {Catinella}, \& {Cortese}}]{Dave20}
{Dav{\'e}}, R., {Crain}, R.~A., {Stevens}, A. R.~H., {et~al.} 2020, \mnras,
  497, 146, \dodoi{10.1093/mnras/staa1894}

\bibitem[{{D{\'e}nes} {et~al.}(2014){D{\'e}nes}, {Kilborn}, \&
  {Koribalski}}]{Denes14}
{D{\'e}nes}, H., {Kilborn}, V.~A., \& {Koribalski}, B.~S. 2014, MNRAS, 444, 667

\bibitem[{{Haynes} {et~al.}(2018){Haynes}, {Giovanelli}, {Kent}, {Adams},
  {Balonek}, {Craig}, {Fertig}, {Finn}, {Giovanardi}, {Hallenbeck}, {Hess},
  {Hoffman}, {Huang}, {Jones}, {Koopmann}, {Kornreich}, {Leisman}, {Miller},
  {Moorman}, {O'Connor}, {O'Donoghue}, {Papastergis}, {Troischt}, {Stark}, \&
  {Xiao}}]{Haynes18}
{Haynes}, M.~P., {Giovanelli}, R., {Kent}, B.~R., {et~al.} 2018, \apj, 861, 49,
  \dodoi{10.3847/1538-4357/aac956}

\bibitem[{{Hoppmann} {et~al.}(2015){Hoppmann}, {Staveley-Smith}, {Freudling},
  {Zwaan}, {Minchin}, \& {Calabretta}}]{Hoppmann15}
{Hoppmann}, L., {Staveley-Smith}, L., {Freudling}, W., {et~al.} 2015, \mnras,
  452, 3726, \dodoi{10.1093/mnras/stv1084}

\bibitem[{{Jones} {et~al.}(2018){Jones}, {Haynes}, {Giovanelli}, \&
  {Moorman}}]{Jones18}
{Jones}, M.~G., {Haynes}, M.~P., {Giovanelli}, R., \& {Moorman}, C. 2018,
  \mnras, 477, 2, \dodoi{10.1093/mnras/sty521}

\bibitem[{{L{\'o}pez-Sanjuan} {et~al.}(2017){L{\'o}pez-Sanjuan}, {Tempel},
  {Ben{\'\i}tez}, {Molino}, {Viironen}, {D{\'\i}az-Garc{\'\i}a},
  {Fern{\'a}ndez-Soto}, {Santos}, {Varela}, {Cenarro}, {Moles}, {Arnalte-Mur},
  {Ascaso}, {Montero-Dorta}, {Povi{\'c}}, {Mart{\'\i}nez}, {Nieves-Seoane},
  {Stefanon}, {Hurtado-Gil}, {M{\'a}rquez}, {Perea}, {Aguerri}, {Alfaro},
  {Aparicio-Villegas}, {Broadhurst}, {Cabrera-Ca{\~n}o}, {Castander}, {Cepa},
  {Cervi{\~n}o}, {Crist{\'o}bal-Hornillos}, {Gonz{\'a}lez Delgado}, {Husillos},
  {Infante}, {Masegosa}, {del Olmo}, {Prada}, \& {Quintana}}]{LopezSanjuan17}
{L{\'o}pez-Sanjuan}, C., {Tempel}, E., {Ben{\'\i}tez}, N., {et~al.} 2017, \aap,
  599, A62, \dodoi{10.1051/0004-6361/201629517}

\bibitem[{{Newman} {et~al.}(2013){Newman}, {Cooper}, {Davis}, {Faber}, {Coil},
  {Guhathakurta}, {Koo}, {Phillips}, {Conroy}, {Dutton}, {Finkbeiner}, {Gerke},
  {Rosario}, {Weiner}, {Willmer}, {Yan}, {Harker}, {Kassin}, {Konidaris},
  {Lai}, {Madgwick}, {Noeske}, {Wirth}, {Connolly}, {Kaiser}, {Kirby},
  {Lemaux}, {Lin}, {Lotz}, {Luppino}, {Marinoni}, {Matthews}, {Metevier}, \&
  {Schiavon}}]{Newman13}
{Newman}, J.~A., {Cooper}, M.~C., {Davis}, M., {et~al.} 2013, \apjs, 208, 5,
  \dodoi{10.1088/0067-0049/208/1/5}

\bibitem[{{P{\'e}roux} \& {Howk}(2020)}]{Peroux20}
{P{\'e}roux}, C., \& {Howk}, J.~C. 2020, \araa, 58, 363,
  \dodoi{10.1146/annurev-astro-021820-120014}

\bibitem[{{Pillepich} {et~al.}(2019){Pillepich}, {Nelson}, {Springel},
  {Pakmor}, {Torrey}, {Weinberger}, {Vogelsberger}, {Marinacci}, {Genel}, {van
  der Wel}, \& {Hernquist}}]{Pillepich19}
{Pillepich}, A., {Nelson}, D., {Springel}, V., {et~al.} 2019, \mnras, 490,
  3196, \dodoi{10.1093/mnras/stz2338}

\bibitem[{{Ponomareva} {et~al.}(2023){Ponomareva}, {Jarvis}, {Pan}, {Maddox},
  {Jones}, {Frank}, {Rajohnson}, {Mulaudzi}, {Meyer}, {Adams}, {Baes}, {Hess},
  {Kurapati}, {Prandoni}, {Sinigaglia}, {Spekkens}, {Tudorache}, {Heywood},
  {Collier}, \& {Sekhar}}]{Ponomareva23}
{Ponomareva}, A.~A., {Jarvis}, M.~J., {Pan}, H., {et~al.} 2023, \mnras, 522,
  5308, \dodoi{10.1093/mnras/stad1249}

\bibitem[{{Rao} \& {Briggs}(1993)}]{Rao93}
{Rao}, S., \& {Briggs}, F. 1993, \apj, 419, 515, \dodoi{10.1086/173504}

\bibitem[{{Rosenberg} \& {Schneider}(2002)}]{Rosenberg02}
{Rosenberg}, J.~L., \& {Schneider}, S.~E. 2002, \apj, 567, 247,
  \dodoi{10.1086/338377}

\bibitem[{{Schaye} {et~al.}(2015){Schaye}, {Crain}, {Bower}, {Furlong},
  {Schaller}, {Theuns}, {Dalla Vecchia}, {Frenk}, {McCarthy}, {Helly},
  {Jenkins}, {Rosas-Guevara}, {White}, {Baes}, {Booth}, {Camps}, {Navarro},
  {Qu}, {Rahmati}, {Sawala}, {Thomas}, \& {Trayford}}]{Schaye15}
{Schaye}, J., {Crain}, R.~A., {Bower}, R.~G., {et~al.} 2015, \mnras, 446, 521,
  \dodoi{10.1093/mnras/stu2058}

\bibitem[{{Willmer} {et~al.}(2006){Willmer}, {Faber}, {Koo}, {Weiner},
  {Newman}, {Coil}, {Connolly}, {Conroy}, {Cooper}, \& {Davis}}]{Willmer06}
{Willmer}, C.~N.~A., {Faber}, S.~M., {Koo}, D.~C., {et~al.} 2006, \apj, 647,
  853, \dodoi{10.1086/505455}

\bibitem[{{Zwaan}(2000)}]{Zwaan00}
{Zwaan}, M.~A. 2000, PhD thesis, Ph.D. Thesis, Groningen: Rijksuniversiteit,
  2000

\bibitem[{{Zwaan} {et~al.}(2001){Zwaan}, {Briggs}, \& {Sprayberry}}]{Zwaan01}
{Zwaan}, M.~A., {Briggs}, F.~H., \& {Sprayberry}, D. 2001, \mnras, 327, 1249,
  \dodoi{10.1046/j.1365-8711.2001.04844.x}

\bibitem[{{Zwaan} {et~al.}(1997){Zwaan}, {Briggs}, {Sprayberry}, \&
  {Sorar}}]{Zwaan97}
{Zwaan}, M.~A., {Briggs}, F.~H., {Sprayberry}, D., \& {Sorar}, E. 1997, \apj,
  490, 173, \dodoi{10.1086/304872}

\bibitem[{{Zwaan} {et~al.}(2005){Zwaan}, {Meyer}, {Staveley-Smith}, \&
  {Webster}}]{Zwaan05}
{Zwaan}, M.~A., {Meyer}, M.~J., {Staveley-Smith}, L., \& {Webster}, R.~L. 2005,
  \mnras, 359, L30, \dodoi{10.1111/j.1745-3933.2005.00029.x}

\end{thebibliography}

\bibliographystyle{aasjournal}	

\appendix
{\section{Fitting a relation to stacked measurements}
\label{appndx:fit}
We assume that the average \hi\ mass of galaxies at $z\approx1$ for a given $\MB$ follows a relation of the following form.

\begin{equation}
\label{eqn:MHI-MB_form}
    \log \MHI= a \, (\MB+21) + c \,,
\end{equation}

We fit the above relation to our measurements of the average $\MHI$ in the three $\MB$ subsamples by taking into account the distribution of $\MB$ values in each subsample as well as the weights assigned to each galaxy. First, for given values of the parameters $(a,c)$, we use  Equation~\ref{eqn:MHI-MB_form} to compute the average \hi\ mass in each of the three $\MB$\ subsamples, $\langle {\MHI}(a,c)\rangle^i$ (for the $i$th $\MB$ subsample). Next, we calculate the $\chi^2$ for the given values of $(a,c)$ as follows:

\begin{equation}
\chi^2 (a,c) = \sum_{i=1}^{3} \left( \frac{\langle{\MHI}\rangle^i-\langle {\MHI}(a,c)\rangle^i}{\sigma^i_{\MHI}}\right)^2 \,,
\label{eqn:chisqr}
\end{equation}
where $\langle{\MHI}\rangle^i$ and $\sigma^i_{\MHI}$  are,  respectively, the measured average \hi\ mass and the RMS uncertainty on the average \hi\ mass for galaxies in the $i$th $\MB$ subsample. We minimize the $\chi^2$ of Equation~\ref{eqn:chisqr} using a steepest-descent algorithm to obtain the best-fit parameters $(a,c)$.

We emphasise that Equation~\ref{eqn:MHI-MB_form} is based on the observed form of the $\MHI-\MB$ relation (and other \hi\ scaling relations) in the local Universe \citep[e.g.][]{Denes14}. Our current measurement of the average \hi\ mass in only three $\MB$ subsamples does not allow for an independent verification of the form of the $\MHI-\MB$ relation at $z\approx1$. This should be possible with deeper \hii\ observations in the future that would yield $\langle\MHI\rangle$ measurements in a larger number of $\MB$ subsamples.
}
\section{Estimating the \hi\ mass function}
\label{appndx}
We estimate the \hi\ mass function by combining the $\MHI-\MB$\ relation and the B-band luminosity function \citep{Briggs90,Rao93,Zwaan01}, appropriately incorporating the effect of the scatter of the $\MHI-\MB$\ relation \citep{Bera22}.

We again assume, based on observational results at $z\approx0$ \citep{Denes14}, that the median $\MHI-\MB$\ relation has the following form,
\begin{equation}
\label{eqn:MHI_MB}
    \log \MHI= a \, \MB + b \,,
\end{equation}
with a log-normal scatter ($\sigma$), such that the probability that a galaxy with a B-band luminosity $\MB$ has a \hi\ mass $\MHI$ is given by
\begin{equation}
    \label{eqn:prob}
    p(\log \MHI | \MB) \ d\MB = \frac{a}{\sqrt{2 \pi} \sigma} \exp{\left[-\frac{\left\{\log \MHI - (a \ \MB + b) \right\}^2}{2 \sigma^2}\right]} d\MB
\end{equation}.

In the hypothetical case of no scatter, i.e. $\sigma=0$, one can simply use Eqn.~\ref{eqn:MHI_MB} and the B-band luminosity function, $\phi_\textrm{B} (\MB)$, to obtain the \hi\ mass function, $\phi_{\sigma=0}(\log \MHI)$, by performing a transformation of variables,
\begin{equation}
    \phi_{\sigma=0} (\log \MHI) d (\log \MHI)  =  \phi_\textrm{B} \left(\frac{\log \MHI-b}{a}\right) \frac{d (\log \MHI)}{a} \,.  
\end{equation}

This estimate of the \hi\ mass function $\phi_{\sigma=0} (\log \MHI)$ is similar to the early estimates of the \himf\ at $z \approx 0$ \citep{Briggs90,Rao93,Zwaan01}. However, in the case of a non-zero scatter in the $\MHI-\MB$ relation, one must take into account the actual probability, from Eqn.~\ref{eqn:prob}, that a galaxy with B-band luminosity ($\MB$) has an \hi\ mass of $\MHI$ to obtain the \hi\ mass function, via the following equation:
\begin{equation}
\label{eqn:himf}
    \phi (\log \MHI) \ d (\log \MHI) = \left[ \int_{-\infty}^{\infty}  \phi_\textrm{B}(\MB) \ p(\log \MHI|\MB) \ d(\MB) \right] \frac{d (\log \MHI)}{a}  \, .
\end{equation}

We use the measurement of $\phi_\textrm{B}(\MB)$ from the ALHAMBRA survey at $z \approx 1$ \citep{LopezSanjuan17}, our measurement of the median $\MHI-\MB$ relation (Eqn.~\ref{eqn:MHI_MB_median}) at $z \approx 1$, and an assumed scatter of 0.26~dex for the $\MHI-\MB$ relation \citep[i.e. the same as that in the local Universe;][]{Denes14}, to numerically evaluate the convolution of Eqn.~\ref{eqn:himf}; this yields the estimate of the \himf\ at $z\approx1$ shown in Figure~\ref{fig:himf}[A].

The errors on the \himf\ are estimated using Monte Carlo simulations. In each Monte Carlo run, the parameters $a, b$ in the $\MHI-\MB$ relation and the parameters $\phi^*,\MB^*, \alpha$ for $\phi_B$ are randomly drawn from Gaussian probability distributions, with standard deviations set to the estimated $1\sigma$ errors on the individual parameters. 
The randomly-drawn values of the parameters are then used to evaluate Eqn.~\ref{eqn:himf} to obtain independent estimates of the \himf\ at $z\approx1$. We repeat the above procedure $10^5$ times to obtain an estimate of the error on our estimate of the \himf\ at $z\approx1$.
\end{document}